\documentclass{article}

\usepackage{arxiv}

\usepackage[utf8]{inputenc} 
\usepackage[T1]{fontenc}    
\usepackage{hyperref}       
\usepackage{url}            
\usepackage{booktabs}       
\usepackage{amsfonts}       
\usepackage{nicefrac}       
\usepackage{microtype}      
\usepackage{lipsum}		
\usepackage{graphicx}
\usepackage[square, numbers]{natbib}
\usepackage{doi}

\usepackage{subcaption}
\usepackage{float}
\usepackage{xcolor}
\usepackage{rotating}

\title{A comparative analysis of~deep learning models for~lung segmentation on~X-ray images}

\author{ \href{https://orcid.org/0000-0003-2903-6050}{\includegraphics[scale=0.06]{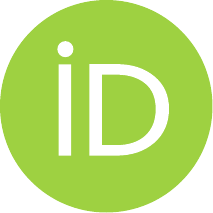}\hspace{1mm}Weronika Hryniewska-Guzik}, \hspace{0.5cm} Jakub Bilski\thanks{equal contribution} \And Bartosz Chrostowski\footnotemark[\value{footnote}], \hspace{0.5cm} Jakub Drak Sbahi\footnotemark[\value{footnote}],\hspace{0.5cm} \href{https://orcid.org/0000-0001-8423-1823}{\includegraphics[scale=0.06]{orcid.pdf}\hspace{1mm}Przemysław Biecek} \vspace{0.5cm}\\
	Faculty of Mathematics and Information Science\\
	Warsaw University of Technology, Poland \vspace{0.2cm}\\ 
	\texttt{weronika.hryniewska.dokt@pw.edu.pl} \\
}

\date{}

\renewcommand{\headeright}{}
\renewcommand{\undertitle}{}
\renewcommand{\shorttitle}{A comparative analysis of~deep learning models for~lung segmentation on~X-ray images}

\hypersetup{
pdftitle={A comparative analysis of deep learning models for lung segmentation on X-ray images},
pdfsubject={q-bio.NC, q-bio.QM},
pdfauthor={Weronika Hryniewska-Guzik, Jakub Bilski, Bartosz Chrostowski, Jakub Drak Sbahi, Przemysław Biecek},
pdfkeywords={Semantic segmentation, X-ray, Lungs, Deep Learning, U-Net},
}

\begin{document}
\maketitle

\begin{abstract}
Robust and highly accurate lung segmentation in X-rays is crucial in medical imaging. This study evaluates deep learning solutions for~this task, ranking existing methods and~analyzing their performance under diverse image modifications. Out of~61 analyzed papers, only nine offered implementation or pre-trained models, enabling assessment of~three prominent methods: Lung VAE, TransResUNet, and~CE-Net. The~analysis revealed that CE-Net performs best, demonstrating the~highest values in dice similarity coefficient and~intersection over union metric. 

\end{abstract}


\section{Introduction}

In the~field of~medical imaging, accurate segmentation of~lungs on~X-rays is important in many applications~\cite{reamaroon2020robust}, from early disease detection to treatment planning and~patient monitoring. As healthcare evolves, the~need for~fast and~accurate tools grows, implying physician support with deep learning approaches~\cite{liu2022automatic}. In particular, solutions such as U-Net demonstrate the~potential to automate the~task of~lung segmentation, offering promising advances in improved accuracy~\cite{unet}.

However, despite these advances, the~inevitable diversity of~X-ray images makes it difficult for~some modern segmentation methods to deal with them. Although many solutions show high performance in simple and~typical cases, their performance degrades when confronted with complex ones. Moreover, the~issue of~using pre-trained models on~images with different characteristics may have potential negative consequences when used for~real-world solutions~\cite{Hinterstoisser_2018_ECCV_Workshops}.

Recognizing these challenges, our objective is to analyze existing solutions for~lung segmentation and~systematically evaluate their performance across a dataset of~varying characteristics. In this study, we analyze and~compare three prominent methods - Lung VAE, TransResUNet, and~CE-Net - using five image modifications. The~ultimate goal is to determine the~most accurate method for~lung segmentation in diverse scenarios.


\section{Methodology}

The complexity of~the~lung segmentation task is related to the~scarcity of~actual data containing X-ray images and~its masks with diverse obstructions such as jewelry, advanced stages of~disease, and~some medical devices present in a~patient's body. To address this limitation, we merged two existing datasets: Montgomery County X-ray~\cite{9} and~Shenzhen Hospital X-ray~\cite{25}. The~first one contains 138 X-rays, of~which 80 are normal and~58 are abnormal with manifestations of~tuberculosis. Alongside the~lung, segmentation is provided. The~second one contains 340 normal and~275 abnormal X-rays showing various manifestations of~tuberculosis. The~masks are provided by Stirenko et al.~\cite{59}.

For our analysis of~models dedicated for~lung field segmentation, we select 54 methods documented in Çallı et al.~\cite{8} and~seven scientific papers which implementation is available on~Github platform models. Most of~the~solutions could not be run due to obsolete versions of~libraries, lack of~the~reproduction steps in the~article/repository, not working parts of~code, and~lack of~methods for~using data other than those provided with code. Finally, we are able to run only three of~them, described in Section \ref{relatedworks}.

If there was a~pre-trained model available for~any of~the~architectures, it was used for~the~evaluation process. Otherwise, a~model was trained on~the~data on~which it was originally trained, as it was described in the~source article. 

\begin{figure}[h]
    \centering
    \includegraphics[width=0.8\textwidth]{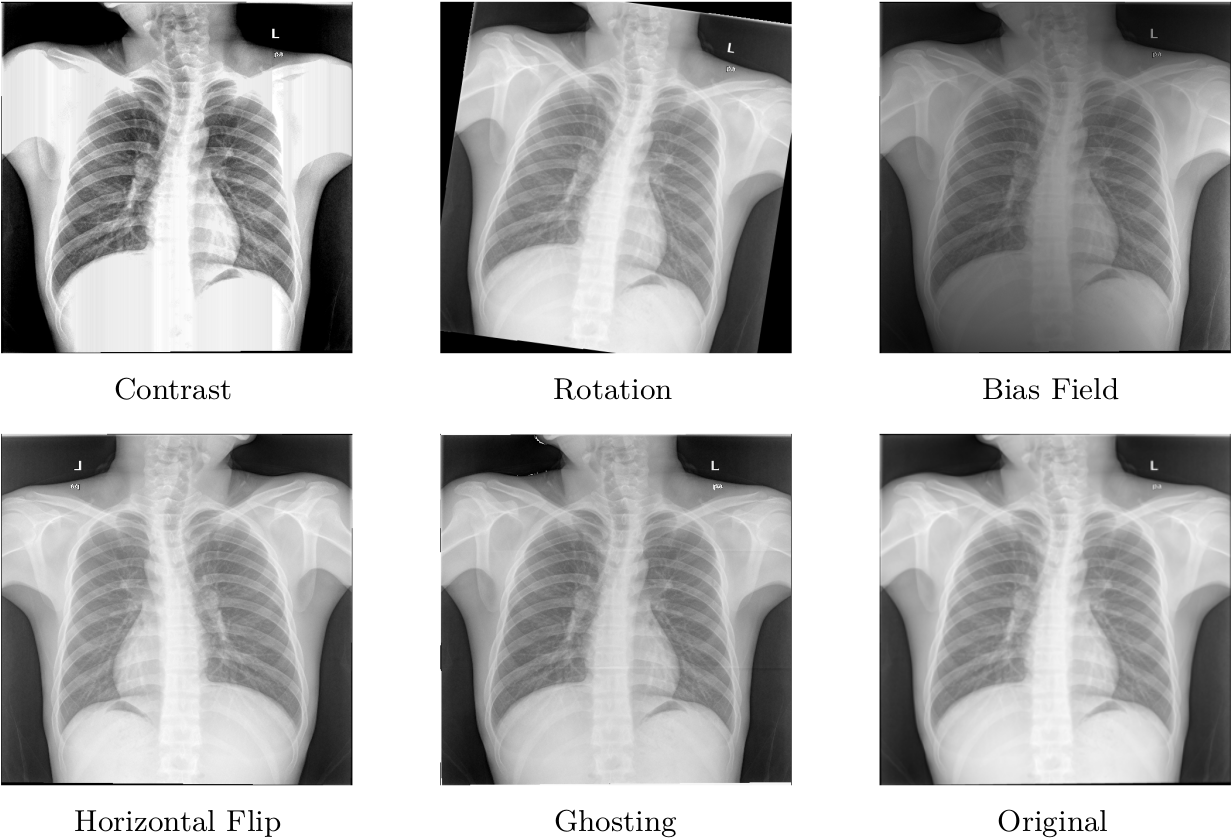}
    \caption{Types of~augmentation on~which the~ability to adapt to new conditions of~segmentation models was tested.}
    \label{fig:augmentation}
\end{figure}

Then, an~evaluation was performed on~the~prepared test data. It contained both original and~augmented images. For every pair of~ground-true and~predicted mask, the~quality of~segmentation was assessed using dice similarity coefficient~\cite{55} and~intersection over union (IoU). The~following augmentations, presented in Figure \ref{fig:augmentation}, were done: contrast, random rotation, bias field, horizontal flip, and~discrete "ghost" artifacts. The~resulting dataset allowed us to perform testing on~model behavior when presented with an~image with augmentation that was not present in the~training phase.


\vspace{-1em}
\section{Related Works}\label{relatedworks}

\textbf{Lung VAE~\cite{53}}\footnote{\url{https://github.com/raghavian/lungVAE}} aim is to be a~lung segmentation model that is not dependent on~training set distribution. For this purpose, a U-net-based segmentation network encoder and~variational encoder is used. As the~authors did not train the~model on~opacity cases, the~variational encoder is used for data imputation and may be treated as data augmentation. The~results of U-net encoder and variational encoder are concatenated and~passed to the~decoder.

\textbf{TransResUNet~\cite{51}}\footnote{\url{https://github.com/sakibreza/TransResUNet}} is an~improved version of~U-net~\cite{unet} model dedicated for~the~task of~lung field segmentation. Three modifications were made. First, a~pre-trained encoder was taken from VGG-16 architecture trained on~the~ImageNet dataset. The~first seven layers were used. As a~second improvement, between the~encoder and~decoder layers were added skip connections with a series of convolution blocks. Finally, a~dedicated post-processing step with hole filling, artefacts removal and morphological opening was applied to~the~output images. 

\textbf{CE-Net~\cite{21}}\footnote{\url{https://github.com/Guzaiwang/CE-Net}} is n~U-net-based approach of~a~context encoder network. Its aim is to capture more abstract and~preserve spatial information for~2D medical image segmentation. There are three components: a~feature encoder module, a~context extractor, and~a~feature decoder module. For feature extraction, pretrained ResNet-34 is used. Dense atrous convolution block and~residual multi-kernel pooling are used for~context extraction. In feature decoder module, transposed convolution is applied.

\section{Results and~discussion}

\begin{figure}[ht]
    \centering
    \rotatebox[origin=l]{90}{\hspace{15mm}dice score}
    \begin{subfigure}{0.26\textwidth}
        \includegraphics[height=49mm]{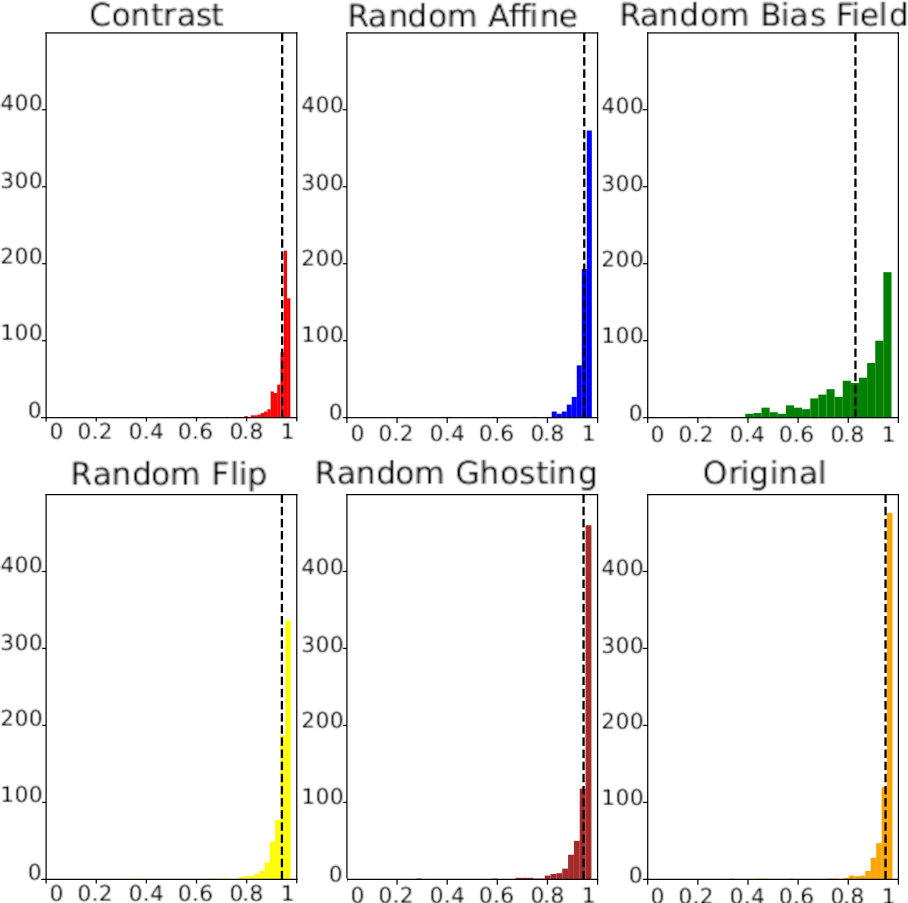}
        \caption{Lung VAE}
        \label{fig:dc_vae}
    \end{subfigure}
    \hfill
    \begin{subfigure}{0.32\textwidth}
        \includegraphics[height=49mm]{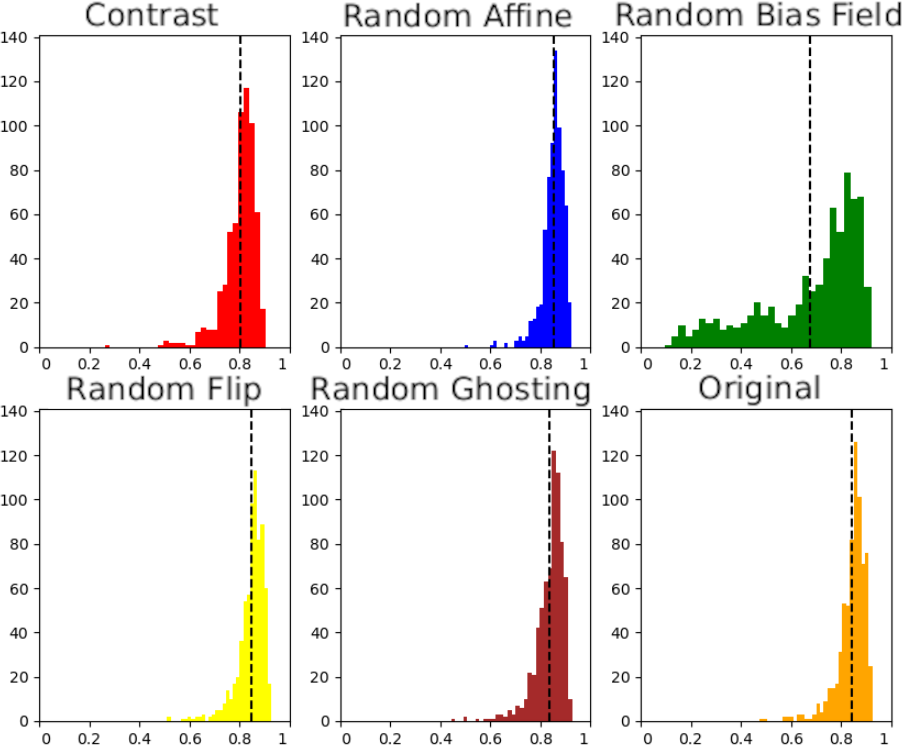}
        \caption{TransResUNet}
        \label{fig:dc_trans}
    \end{subfigure}
    \hfill
    \begin{subfigure}{0.3\textwidth}
        \includegraphics[height=49mm]{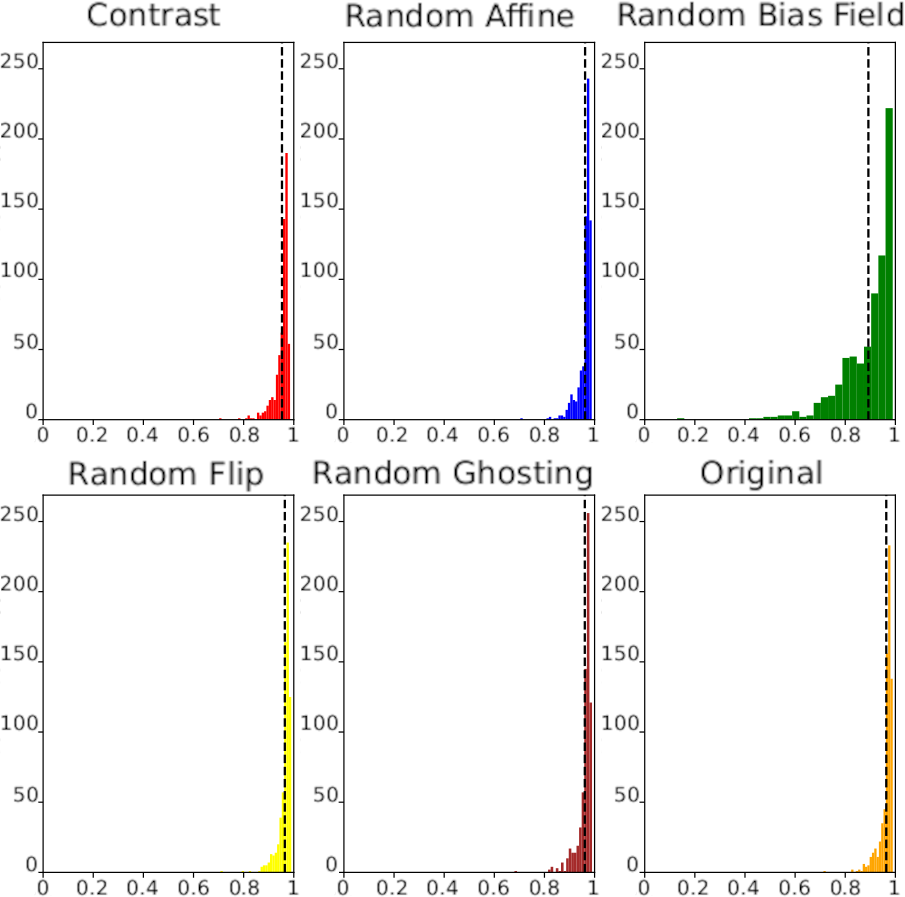}
        \caption{CE-Net}
        \label{fig:dc_cenet}
    \end{subfigure}

    \rotatebox[origin=lB]{90}{\hspace{20mm}IoU}
    \begin{subfigure}{0.26\textwidth}
        \includegraphics[height=49mm]{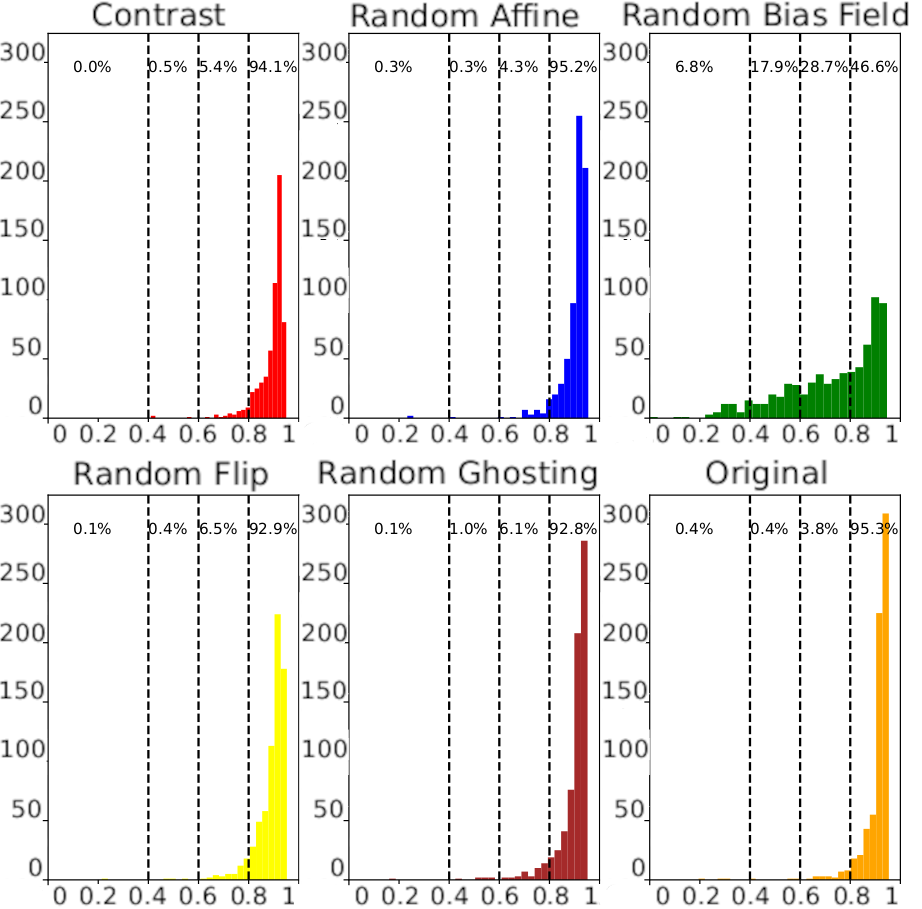}
        \caption{Lung VAE}
        \label{fig:iou_vae}
    \end{subfigure}
    \hfill
    \begin{subfigure}{0.32\textwidth}
        \includegraphics[height=49mm]{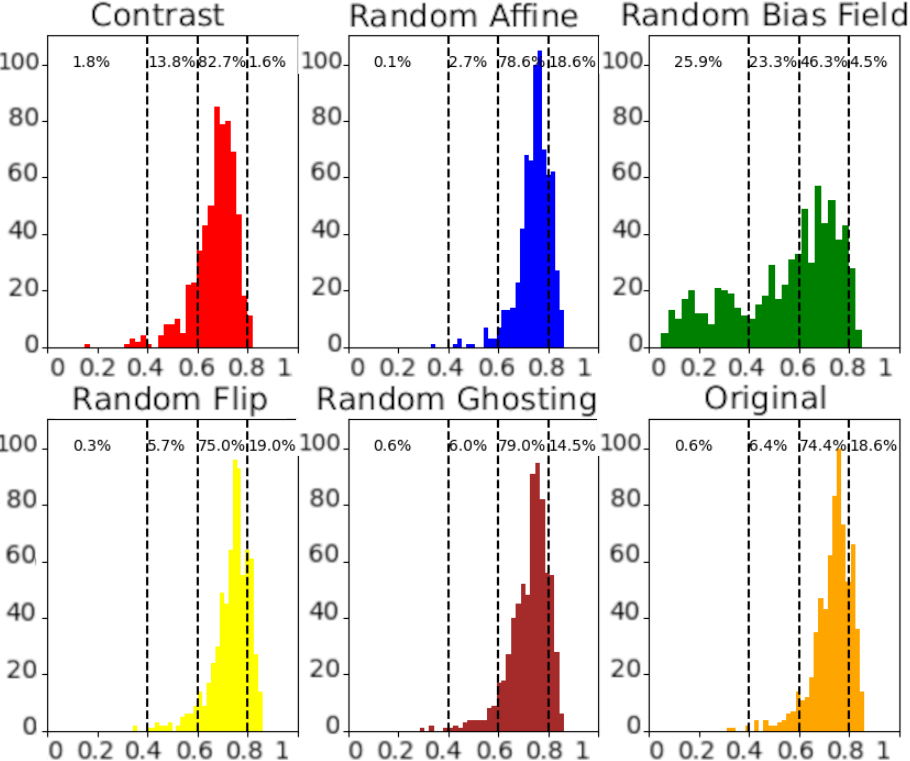}
        \caption{TransResUNet}
        \label{fig:iou_trans}
    \end{subfigure}
    \hfill
    \begin{subfigure}{0.3\textwidth}
        \includegraphics[height=49mm]{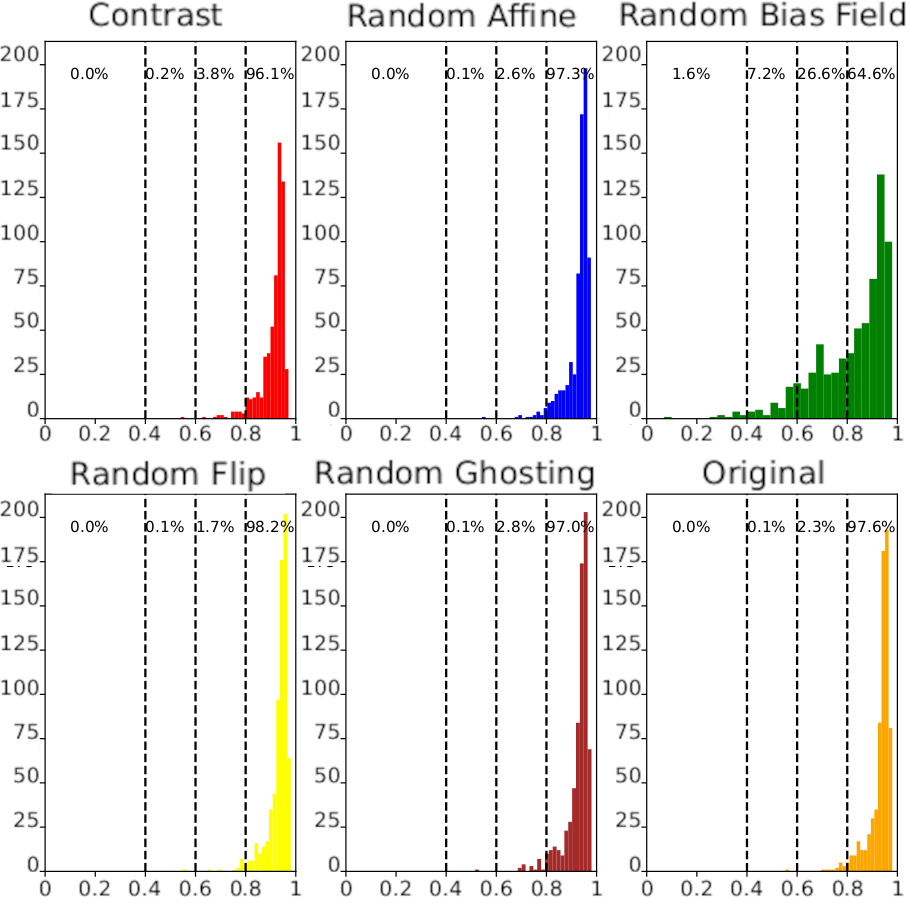}
        \caption{CE-Net}
        \label{fig:iou_cenet}
    \end{subfigure}
    \caption{Segmentation results (dice similarity coefficient and~IoU value) after applying various augmentation methods that have not been performed on~the~training set before. On the~horizontal axis is the~dice similarity coefficient value from 0 to 1, and~on the~vertical axis is the~number of~samples from 0 to 500.}
    \label{fig:dc_iou}
\end{figure}

The Lung VAE model, shown in Figure \ref{fig:dc_vae}, obtained the~best results for~images without augmentations and~the~worst for~the~Random Bias Field augmentation. From the~augmented images, it seems to have the~best results for~the~Random Ghosting.

Figure \ref{fig:dc_trans} illustrates the~results for~the~TransResUNet model, which performed less favorably compared to the~other two methods. The~mean dice loss for~images without augmentation was slightly greater than 80\%. Three augmentation types had no real effect on~the~results; Random Affine, Random Flip, and~Random Ghosting achieved a~very similar mean dice loss as the~case with no augmentation. The~two remaining augmentation methods posed a~more serious problem. Contrast got a~slightly lower dice loss and~a~thicker tail of~the~distribution, while Random Bias Field performed much worse, with a~very long tail and~low mean score. While other methods also struggled with this type of~augmentation, this method seemed to perform the~worst.

The results for~CE-Net are presented in Figure \ref{fig:dc_cenet}, showcasing its outperformance relative to both Lung VAE and~TransResUNet. This distinction is particularly noticeable for~the~Random Bias Field augmentation, where CE-Net exhibited a~better average dice similarity coefficient and~a~significantly shorter low-tail compared to the~other two models.

Figure \ref{fig:iou_vae}, \ref{fig:iou_trans}, and~\ref{fig:iou_cenet} illustrate the~results for~the~IoU score, providing a~summary of~the~score distribution. Here, the~robustness of~CE-Net is evident, with very few samples falling outside the~best score bin. In contrast, TransResUNet displayed a~thicker tail, with some samples consistently reaching the~worst bin and~a~peak in the~third bin. Although Lung VAE performed slightly worse than CE-Net in all cases, it still achieved significantly better results than TransResUNet, emphasizing CE-Net as the~superior and~more reliable algorithm.

\begin{figure}[ht]
    \centering
    \begin{subfigure}{0.31\textwidth}
        \includegraphics[width=\textwidth]{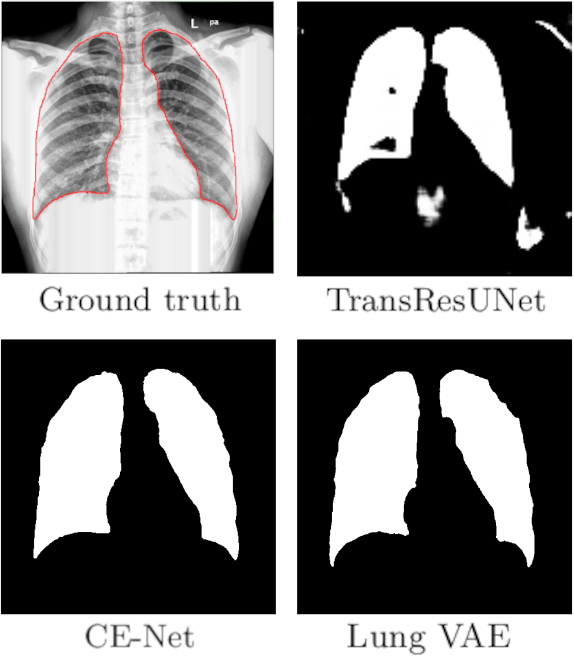}
        \caption{contrast}
        \label{fig:contrast}
    \end{subfigure}
    \hfill
    \begin{subfigure}{0.31\textwidth}
        \includegraphics[width=\textwidth]{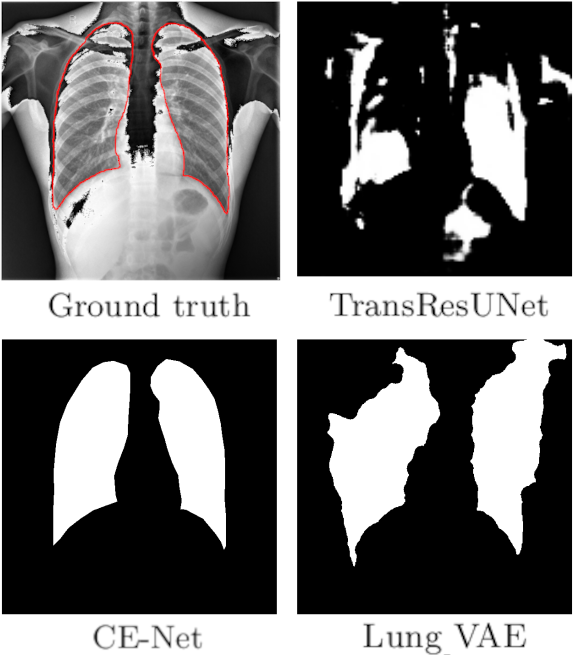}
        \caption{bias field}
        \label{fig:bias_field}
    \end{subfigure}
    \hfill
    \begin{subfigure}{0.3\textwidth}
        \includegraphics[width=\textwidth]{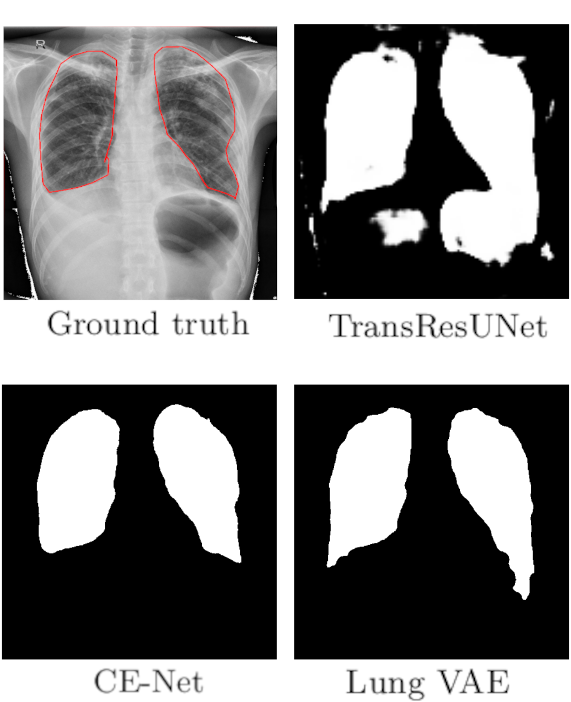}
        \caption{random ghosting}
        \label{fig:ghosting}
    \end{subfigure}
    \caption{Original X-ray images with segmented lungs (marked in red) compared to lung masks generated by various models using different augmentation methods.}
    \label{fig:comparison}
\end{figure}

Figure \ref{fig:comparison} shows an~example of~the~segmentation results of~the~original image for~three models using different previously unseen augmentations. As was presented in the~aggregated analysis, the~CE-Net model in all augmentations managed to obtain satisfying results. The~most significant differences between outputs of~CE-Net and~Lung VAE are visible in Figure \ref{fig:bias_field}. The~Lung VAE model captured the~position of~the~lung. However, the~obtained mask did not manage to preserve the~shape, especially in the~top part of~the~image. TransResUNet often detects spots outside of~the~lung that seem to correspond with places with a~darker shade of~gray than the~rest of~the~body.

\vspace{-1em}
\section{Conclusions}

In literature, the~vast majority of~solutions for~lung segmentation are based on the~U-net architecture enhanced with either the~preprocessing of~data or~the~post-processing of~output masks. It is worth stressing that when proposing a new architecture, it is necessary to ensure that the results are reproducible. Among the~61 examined papers, only three demonstrated effective solutions: CE-Net, TransResUNet, and~Lung VAE. 

Evaluation of these lung segmentation methods based on~dice loss and~IoU metrics revealed consistent superiority of~CE-Net across experiments. Notably, TransResUNet exhibited limitations, struggling to accurately localize the~lungs in certain instances. The~direct comparison of~generated masks further emphasized the~robust performance of~CE-Net over both TransResUNet and~Lung VAE. 

These findings highlight the~challenges in achieving reliable and~consistent results in deep learning for~segmentation tasks, underscoring the~significance of~methodological choices in model development. The code is available at \url{https://github.com/Hryniewska/lung-segmentation-on-X-rays}.

\section*{Acknowledgment}
The~authors would like to thank Jakub Brojacz for his~valuable impact on the research.

\newpage
\bibliographystyle{abbrvnat}
\bibliography{references}

\begin{thebibliography}{12}
\providecommand{\natexlab}[1]{#1}
\providecommand{\url}[1]{\texttt{#1}}
\expandafter\ifx\csname urlstyle\endcsname\relax
  \providecommand{\doi}[1]{doi: #1}\else
  \providecommand{\doi}{doi: \begingroup \urlstyle{rm}\Url}\fi

\bibitem[Candemir et~al.(2014)Candemir, Jaeger, Palaniappan, Musco, Singh, Xue, Karargyris, Antani, Thoma, and McDonald]{9}
S.~Candemir, S.~Jaeger, K.~Palaniappan, J.~P. Musco, R.~K. Singh, Z.~Xue, A.~Karargyris, S.~Antani, G.~Thoma, and C.~J. McDonald.
\newblock Lung segmentation in chest radiographs using anatomical atlases with nonrigid registration.
\newblock \emph{IEEE Transactions on Medical Imaging}, 33\penalty0 (2):\penalty0 577--590, 2014.
\newblock \doi{10.1109/TMI.2013.2290491}.

\bibitem[Gu et~al.(2019)Gu, Cheng, Fu, Zhou, Hao, Zhao, Zhang, Gao, and Liu]{21}
Z.~Gu, J.~Cheng, H.~Fu, K.~Zhou, H.~Hao, Y.~Zhao, T.~Zhang, S.~Gao, and J.~Liu.
\newblock {CE-Net: Context Encoder Network for 2D Medical Image Segmentation}.
\newblock \emph{IEEE Transactions on Medical Imaging}, 38\penalty0 (10):\penalty0 2281--2292, 2019.
\newblock \doi{10.1109/TMI.2019.2903562}.

\bibitem[Hinterstoisser et~al.(2018)Hinterstoisser, Lepetit, Wohlhart, and Konolige]{Hinterstoisser_2018_ECCV_Workshops}
S.~Hinterstoisser, V.~Lepetit, P.~Wohlhart, and K.~Konolige.
\newblock On pre-trained image features and synthetic images for deep learning.
\newblock In \emph{Proceedings of the European Conference on Computer Vision (ECCV) Workshops}, September 2018.

\bibitem[Jaeger et~al.(2014)Jaeger, Karargyris, Candemir, Folio, Siegelman, Callaghan, Xue, Palaniappan, Singh, Antani, Thoma, Wang, Lu, and McDonald]{25}
S.~Jaeger, A.~Karargyris, S.~Candemir, L.~Folio, J.~Siegelman, F.~Callaghan, Z.~Xue, K.~Palaniappan, R.~K. Singh, S.~Antani, G.~Thoma, Y.-X. Wang, P.-X. Lu, and C.~J. McDonald.
\newblock Automatic tuberculosis screening using chest radiographs.
\newblock \emph{IEEE Transactions on Medical Imaging}, 33\penalty0 (2):\penalty0 233--245, 2014.
\newblock \doi{10.1109/TMI.2013.2284099}.

\bibitem[Liu et~al.(2022)Liu, Luo, and Yang]{liu2022automatic}
W.~Liu, J.~Luo, and e.~a. Yang, Yu.
\newblock Automatic lung segmentation in chest x-ray images using improved u-net.
\newblock \emph{Scientific Reports}, 12:\penalty0 8649, 2022.
\newblock \doi{10.1038/s41598-022-12743-y}.

\bibitem[Reamaroon et~al.(2020)Reamaroon, Sjoding, and Derksen]{reamaroon2020robust}
N.~Reamaroon, M.~W. Sjoding, and e.~a. Derksen, Hanneke.
\newblock Robust segmentation of lung in chest x-ray: Applications in analysis of acute respiratory distress syndrome.
\newblock \emph{BMC Medical Imaging}, 20\penalty0 (1):\penalty0 116, 2020.
\newblock \doi{10.1186/s12880-020-00514-y}.

\bibitem[Reza et~al.(2020)Reza, Amin, and Hashem]{51}
S.~Reza, O.~B. Amin, and M.~Hashem.
\newblock {TransResUNet: Improving U-Net Architecture for Robust Lungs Segmentation in Chest X-rays}.
\newblock In \emph{TENSYMP}, pages 1592--1595, 2020.
\newblock \doi{10.1109/TENSYMP50017.2020.9230835}.

\bibitem[Ronneberger et~al.(2015)Ronneberger, Fischer, and Brox]{unet}
O.~Ronneberger, P.~Fischer, and T.~Brox.
\newblock {U-Net}: Convolutional networks for biomedical image segmentation.
\newblock In N.~Navab, J.~Hornegger, W.~M. Wells, and A.~F. Frangi, editors, \emph{MICCAI}, pages 234--241, 2015.
\newblock \doi{10.1007/978-3-319-24574-4_28}.

\bibitem[Selvan et~al.(2020)Selvan, Dam, Detlefsen, Rischel, Sheng, Nielsen, and Pai]{53}
R.~Selvan, E.~B. Dam, N.~S. Detlefsen, S.~Rischel, K.~Sheng, M.~Nielsen, and A.~Pai.
\newblock Lung segmentation from chest {X-rays} using variational data imputation.
\newblock \emph{Artemiss workshop at ICML}, 2020.

\bibitem[Shamir et~al.(2019)Shamir, Duchin, Kim, Sapiro, and Harel]{55}
R.~R. Shamir, Y.~Duchin, J.~Kim, G.~Sapiro, and N.~Harel.
\newblock Continuous dice coefficient: a method for evaluating probabilistic segmentations.
\newblock \emph{arxiv}, 2019.

\bibitem[Stirenko et~al.(2018)Stirenko, Kochura, Alienin, Rokovyi, Gordienko, Gang, and Zeng]{59}
S.~Stirenko, Y.~Kochura, O.~Alienin, O.~Rokovyi, Y.~Gordienko, P.~Gang, and W.~Zeng.
\newblock Chest {X-Ray} analysis of tuberculosis by deep learning with segmentation and augmentation.
\newblock In \emph{IEEE ELNANO}, Apr. 2018.
\newblock \doi{10.1109/elnano.2018.8477564}.

\bibitem[Çallı et~al.(2021)Çallı, Sogancioglu, {van Ginneken}, {van Leeuwen}, and Murphy]{8}
E.~Çallı, E.~Sogancioglu, B.~{van Ginneken}, K.~G. {van Leeuwen}, and K.~Murphy.
\newblock Deep learning for chest {X-ray} analysis: A survey.
\newblock \emph{Medical Image Analysis}, 72:\penalty0 102125, 2021.
\newblock \doi{10.1016/j.media.2021.102125}.

\end{thebibliography}

\end{document}